\documentclass[preprint,prc,aps,nofootinbib]{revtex4-1}
\newcommand{\be}{\begin{eqnarray}}
\newcommand{\ee}{\end{eqnarray}}

\newcommand{\ave}[1]{\left\langle #1 \right\rangle}
\newcommand{\abs}[1]{\left| #1 \right|}
\newcommand{\eqcomma}{\phantom{A},\phantom{A}}
\newcommand{\etaprime}{\eta^{\prime}}
\usepackage{epsfig}
\begin{document}
\title{ $\etaprime$ Production in Nucleus-Nucleus collisions as a probe of chiral dynamics}
\author{Giorgio Torrieri}
\affiliation{IFGW, State University of Campinas, Brazil}
\email{torrieri@ifi.unicamp.br}
 \begin{abstract}
   I argue that, because of the peculiar properties of the $\etaprime$ meson, it is a promising probe of ``chiral'' dynamics.  In particular, I show that a rotating gluon-dominated plasma might lead to an enhanced production of $\etaprime$ w.r.t. statistical model expectations.   The presence of a strong topological susceptibility might give a similar effect.   In both cases, unlike the statistical model,I expect a non-trivial dependence on event geometry, such as initial volume and impact parameter.
   Hence, an observation of $\etaprime/\pi$ ratio depending strongly on impact parameter might be a good indication of chiral effects, either from vorticity or topological phases of QCD.
 \end{abstract}
\maketitle
 \section{Introduction}
 A recent topic which has been subject of a lot of theoretical and experimental investigation is chiral dynamics, a class of phenomena involving the transfer of angular momentum between ``collective'' angular momentum and spin, in the context of the strongly coupled system created in heavy ion collisions.

 Two main effects are thought to be taking place:  One is the temporary breaking of CP symmetry, in a randomly fluctuating direction, due to topologically non-trivial configurations.   The other, CP-conserving, is the transfer of angular momentum between vorticity and spin due to near-equilibrium statistical mechanics.   The phenomenological manifestations of this are known as Chiral Magnetic and vortaic effects, as well as Global polarization \cite{magnetic0,magnetic1,taskforce,spinpol,surowka,stephanov,gt1,gt2,flork1,flork2,uspol,wangpol,becattini1}.  The two effects are very different (one anomalously breaks the symmetries of QCD, the other does not and should appear in the ideal hydrodynamic limit \cite{gt1,gt2}), but the experimental signatures developed so far are generally insensitive to this distinction \cite{teryaev}.

Indeed, the immediate problem with these observables is that heavy ion events are messy both from a theoretical and experimental perspectives.   The experimental probes of these effects are typically very complicated and subject to production by  ``background`` non-chiral processes \cite{magnetic1,taskforce,spinpol}.
The theory is typically strongly coupled and opaque to effective expansions, making reliable quantitative calculations difficult \cite{surowka,stephanov,gt1,gt2}.

In this work, I would like to propose an seemingly indirect but promising observable as a probe for the presence of this type of effects.   While it is not ''trivial'' to measure, it is a chemical abundance, and hence might be less susceptible to hydrodynamic background effects plaguing other observables:  The momentum-integrated abundance of the $\etaprime$ resonance, reconstructed by peak identification, relative to particles of similar composition (experimentally the best choice is the $\pi^0$), and the dependence of this relative abundance on impact parameter.

The $\etaprime$ meson is considerably heavier than similar pseudo-scalar states,
such as the $\pi, K$ and $\eta$.   The reason is believed to be that the $\etaprime$
aquires mass via the presence of non-perturbative objects such as 
instantons in the QCD vacuum, and consequent breaking of $U_A(1)$ symmetry by these objects, which contribute to the triangle 
anomaly \cite{th,wit,forkel}.

 \begin{figure*}
   \begin{center}
\epsfig{width=1.\textwidth,figure=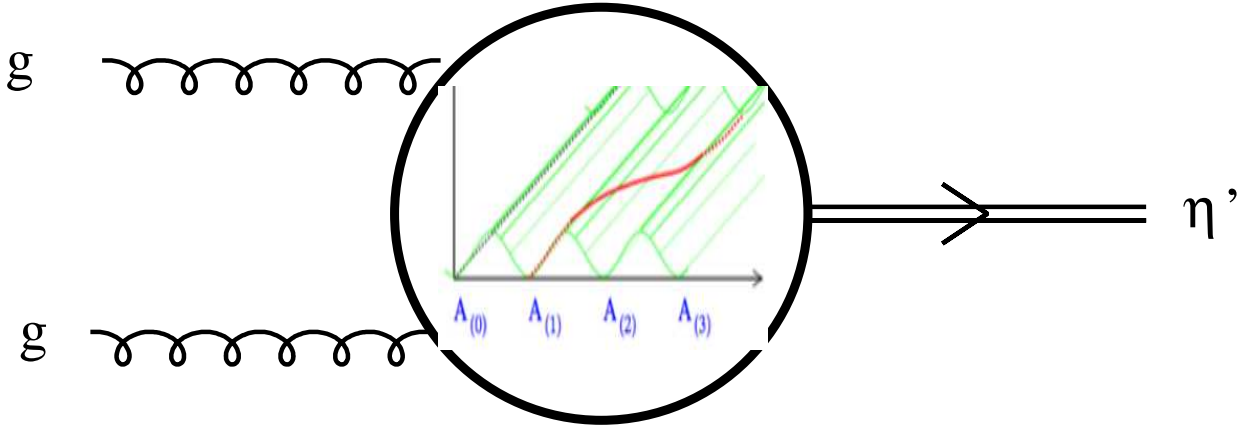}
\caption{\label{figdiag} The production mechanism of the $\etaprime$ from ``free'' gluons via a topologically non-trivial field configuration
  }
\end{center}
\end{figure*}
 If this is true, this anomaly also allows for a new mechanism for generating $\etaprime$ from gluon-gluon fusion, with free gluons fusing into an $\etaprime$ via a topologically non-trivial field configuration (Fig. \ref{figdiag}).

 From symmetry considerations the 
 effective vertex for gluon-gluon-$\etaprime$ must be \cite{as}
 \be
 T^{\alpha\beta}_{ab}(p,q,P)
 = H_f(p^2, q^2, P^2)\, \delta_{ab} \,
 P(p,q,\alpha,\beta)
 \label{eq:vertex}
 \ee
where $p$ and $q$ are the momanta of the gluons, $P$ is the momentum 
 of the produced $\etaprime$ and
 \be
 \label{vertex2}
 P(p,q,\alpha,\beta)= \epsilon_{\mu\nu\lambda\gamma}\, 
 p^\mu \, q^\nu \,
 \epsilon^{\lambda}(p,\alpha) \,
 \epsilon^{\gamma}(q,\beta)
 \ee
Here, $\epsilon_{\lambda}(p,\alpha)$ is the
polarization vector of a gluon with momentum $p$ and helicity 
$\alpha=\pm 1$.  The Kronecker delta $\delta_{ab}$ picks 
the color-neutral combination of the gluons.
The hadronic form factor $H_f(p^2, q^2, P^2)$, as usual, gives the momentum structure of the hadron in terms of the quark wavefunctions.   The corresponding cross-section is
\be
d\hat{\sigma}^{gg\rightarrow \etaprime}_{ab,\alpha\beta}
=  M \delta^4(P-p-q)
{d^3 P \over (2\pi)^3 2 P^0} \eqcomma M={1 \over 4 \sqrt{(p_\mu  q^\mu)^2}}
|T_{ab}^{\alpha\beta}|^2 (2\pi)^4
\label{eq:dsig}
\ee
This was used in \cite{sangyong,sangyong2} to propose the $\etaprime$ as a probe of the double spin asymmetry in the parton distribution function, since the above expressions would have lead to a finite difference $\Delta \sigma^{gg \etaprime} =d\sigma_{++} - d\sigma_{+-} \ne 0$
where $ d\sigma_{++}$ denotes the cross section where both protons have their
spins parallel to their momenta while  $ d\sigma_{+-}$ denotes the
cross section when one proton has its spin anti-parallel to its momentum.
The difference would be directly dependent on the polarized part of the parton distribution function $\Delta G$ given the usual deep inelastic scattering cross-sections
\[\ 
d\Delta\sigma^{pp\rightarrow \etaprime X} = \int dx_1 dx_2 \Delta G(x_1,Q_f^2)  
\Delta G(x_2,Q_f^2)  d\Delta \sigma^{gg\rightarrow \etaprime}
\]
Thus, the form of Eq. \ref{eq:vertex} directly leads to the link between the absolute $\etaprime$ meson abundance and the presence of ``chiral'' QCD fields at the quark level.   This is at first confusing, since $\etaprime$ is a spin (pseudo)scalar state, but the matrix element of Eq. \ref{eq:vertex} ensures that any ``momentum vorticity'' in a gluon-rich medium leads to an enhanced production of $\etaprime$.  Such momentum vorticity is exactly what most chiral processes examined in the literature lead to.

I shall assume the form of Eq. \ref{eq:vertex} has analogues during hadronization of the quark gluon plasma and show that the abundance of the $\etaprime$ meson can function as a probe of chiral dynamics in this context.

I note that the measurement of the $\etaprime$ to study parity-odd bubbles has been proposed long ago \cite{magnetic0,parity1,parity2,referef1,referef2,referefexp1,referefexp2,referefexp2.5,referef3,referef4}. Most of the proposed masurements had to do with the experimentally difficult to detect direct interaction of the $\etaprime$ field  with the modified chiral condensate, leading to meson mass shifts, $\etaprime -\eta$ mixings, and new decay modes.

Direct measurements of the cross-section of $\etaprime$ photo-production in nuclear matter \cite{referefexp3} do offer evidence of vacuum-driven changes in cross-section.  However, even medium-modified $\etaprime$ might be too long lived to experience a mass and width shift observable in experiment, due to the $\etaprime$ large lifetime and weak interaction with the medium.
In this case, the in-medium $\etaprime$ spectral function would relax to its vacuum value before decay.
Other proposals \cite{referefexp1}  bypass this difficulty by an indirect measurement of the $\etaprime$ abundance via $\pi-\pi$ momentum space correlations. many dynamical effects, from flow to Coloumb corrections, are likely to affect these correlations, but past searches in this direction seem to suggest
enhanced production and in-medium mass modification do occur \cite{referefexp2,referefexp2.5}.

Given these encouraging indications I argue that, given the subsequent phenomenological success of the statistical model in decribing momentum-integrated particle ratios \cite{jansbook,share,pbm,becastat},the lack of a conclusive signature for resonance mass shifts \cite{na60,starres} and no solid evidence for a long hadronic phase a simple chemical ratio centrality dependence might be enough, both in the presence of a vorticose medium and of topologically non-trivial regions.   In the next two sections I will show how each of these mechanisms will result in a centrality-dependent enhancement, w.r.t. statistical model expectations, of ratios such as $\etaprime/\pi$ and $\etaprime/\phi$.

\section{$\etaprime$ From vortical hydrodynamics \label{vortical}}
I shall adopt the arguments of the previous section, and apply the concepts behind Eq. \ref{eq:vertex} and Eq. \ref{vertex2} to a rotating thermal quark-gluon plasma.
To do this, one must treat Eq. \ref{eq:vertex} as a kernel for thermal gluons to coalesce into $\etaprime$ at hadronization, thereby neglecting the loss term \cite{sangyong2}.   Because I am interested in an estimate, I shall make some simplifying assumptions such as neglecting mass and coupling temperature dependence computed carefully in \cite{sangyong2}.
The number of $\etaprime$ will then be given by 
 \be
n(P)
 = \sum_{\alpha,\beta} \int \frac{d^3 p}{p^0} \frac{d^3q}{q^0}  M(p,q) \int d^3x_p d^3 x_q \, \delta_{ab} \, 
f_g(x_p,p) f_g(x_q,q) 
 \label{thermal}
 \ee
 here, $f_g(x,p)$ is the gluon distribution in phase space, assumed to be a Juttner-type distributions controlled by a temperature field $T(x)$ and a velocity field $u_\mu(x)$ given by a hydrodynamic simulation.  Up to normalization
  \be
\label{juttner}
  f_g(x,p) \sim T^{-3} \exp \left[ \frac{u_\mu(x) p^\mu}{T(x)}  \right]
  \ee
In the context of Eq. \ref{eq:vertex} and deep inelastic scattering, $H_f$ was a form-factor.  However, there is no reason it cannot be straight-forwardly generalized to a different enviroenment.   Here, therefore, I will use
     $H(...)$ analogously to the previous section, but treat it as a Wigner function. In particular, just as with form factors, it is reasonable to assume it can be expressed in position or momentum space
\begin{equation}
  H(p, q, P^2) = \int \exp\left[ i \left( p_\mu  x_p^\mu + q_\mu x_q^\mu \right) \tilde{H}(x_p,x_q, P^2) \right] d^3 x_p d^3 x_q
\end{equation}
For a qualitative estimate, one assumes that gluons in a comoving frame are approximately unpolarized, i.e. the direct transfer of vorticity to spin studied in \cite{gt1,gt2} is a small effect.
Now, since thermal fluctuations cancel out one has that 
\begin{equation}
  \int \left[  \frac{d^3p}{p^0} \right] f(x,p) p^\mu  \propto T u^\mu(x)
  \end{equation}
further assuming $H(...)$ to be ''small'' w.r.t. the variation in $u_\mu$
\begin{equation}
 | \tilde{H}(x_p, x_q) |^2 \sim \Theta\left[ \left( x_p-x_q \right)^2 -\Delta^2 \right]
\end{equation}
One can directly relate the $\etaprime$ production from such ``anomalous coalescence'' to the vorticity
 \be
\Delta n(P) \sim T^2 \Delta^3 \abs{\Omega}^2
 \label{etavort}
 \ee
 where $\Omega_{\alpha \beta} = \epsilon_{\alpha \beta \gamma \mu} \partial^\gamma u^\mu$ is the vorticity.

 ``Normal'' particles, whose production is not straight-forwardly related to instantons, will not feel any effect from vorticity and their abundance will simply depend, in either a statistical model or coalescence from a thermal medium, on mass, temperature and hadronization volume $V$ \cite{jansbook,share,pbm,becastat}.   For systems where the freezeout is fast enough, and the volume is large enough that dynamical Bose-Einstein correlations are negligible one gets that \cite{jansbook} \footnote{This assumption is not physically guaranteed.  For example, it would not apply to ``halo'' dominated models \cite{halo} where freezeout timescales are comparable with expansion.   In this work I calculate $p_T$ integrated abundances, where the $\pi$ seems to not have centrality-dependent deviations from the statical model \cite{pi0alice}, so assuming these effects are small is justifiable.  To check their size, the dependence of the $\etaprime/\phi$ as well as the $\etaprime/\pi$ ratio might be used, since the $\phi$ is less susceptible to these effects  }
 \[\  n_{m \ll T } \sim V T^3 \eqcomma n_{m \geq T} \sim V T m^2 K_2 \left(\frac{m}{T} \right)\] 
  Hence, relating the $\etaprime$ to another particle, say $\etaprime/\pi$, a clear effect should be seen in more vorticose events.
 The thermal production of $\pi^0$ is completely insensitive to flow and will be, in the massless limit, simply $\sim VT^3$ where $V$ is the total volume. 
 \begin{equation}
  \frac{\etaprime}{\pi} \sim \frac{ \Delta^3 \sum_j \abs{\Omega_j}^2}{V T}+ \left. \frac{\etaprime}{\pi} \right|_{thermal} 
 \end{equation}
 Where $\Omega_i$ is the vorticity associated with an fluid cell $j$ of size $\Delta^3$, summed over the whole event and the thermal background, $\left. \frac{\etaprime}{\pi} \right|_{thermal} \sim \exp\left[ \frac{-m_\etaprime + m_\pi}{T} \right] $ is expected to only depend on the freeze-out temperature $T$ and, in particular, be independent of the production volume.
 
 In other words, there would be a direct dependence of $\etaprime/\pi$ on the impact parameter (as opposed to multiplicity, like strangeness), something that can be seen in scans with system size or in event engineering.
This estimate is shown in Fig. \ref{figeta} (right panel (b)) for top RHIC energies, assuming a $\Delta=1fm$ and a $T=170$ MeV.   The angular momentum and volume are taken from Fig. 3 of \cite{becapol}.   For the $\pi$ I use the statistical model implemented in SHARE \cite{share}.   As an estimate of the statistical model ''background'' I also include (Fig. \ref{figeta} left panel) an estimate, as a function of temperature, of the statistical model value of this ratio, which is of course constant with volume.  The relation between centrality and number of participants was further estimated using the formulae in \cite{florkcent}.
 It should be reiterated that this is an illustrative estimate, and it is the impact parameter dependence, rather than the ratio's absolute value, that will provide the experimental probe.

 In this respect, it is important to underline that I expect $\etaprime/\pi$ yield due to the mechanism described in this section to increase strongly with decreasing centrality.
 This dependence is both physically obvious and markedly different from other mechanisms, where particle ratios are typically independent of centrality or, as in the case of strangeness, rise with centraliy \cite{davidn}.  
 Thus, a strong $\etaprime/\pi$ as a function of impact parameter could signal vorticity at the level of a gluon rich quark-gluon plasma.
 \section{$\etaprime$ From topological charge fluctuations\label{cme} }
In a topologically non-trivial thermal QCD environment \cite{magnetic1}  the Hamiltonian acquires an effective term
\begin{equation}
  \label{deltah}
 \Delta H = \hat{\theta} F_{\mu \nu} \tilde{F}^{\mu \nu} =   \hat{\theta} \epsilon_{\mu \nu \alpha \beta} F^{\mu \nu} \tilde{F}^{\alpha \beta} = \hat{\theta} \epsilon_{\mu \nu \alpha \beta} \epsilon^{\mu \nu \zeta \iota}  \epsilon^{\alpha \beta \rho \omega} \partial_{\zeta} A_{ \iota}  \partial_{\rho} A_{ \omega} 
 \end{equation}
 where $\hat{\theta}$ is the ``random'' value of the effective $\theta$ angle, approximately Gaussian centered around zero and with width related to the topological susceptibility $\ave{\theta^2}$. Explaining the fact that the distribution of $\hat{\theta}$ is centered around zero is known as the ``strong CP problem'', but in a topologically non-trivial medium $\hat{\theta}$ should fluctuate randomly in spacetime, something seen in lattice calculations \cite{sharma,fodor}. 

 As discussed in the introduction, the fact that $\etaprime$ production is closely associated with this topological term has been noted a long time ago and explored with a variety of techniques involving spectral functions and particle correlations.
 Here, I wish to take the simple approach argued for in the introduction and show that, even assuming that the $\etaprime$ is produced, and decays un-modified from its vacuum state by a coalescence type process, topologically non-trivial states will still leave an imprint in the momentum-integrated $\etaprime/\pi$ ratio as a function of multiplicity which can be looked for in an invariant mass $\etaprime \rightarrow \gamma \gamma$ reconstruction and consequent $\etaprime/\pi$ measurement.  
 
 I shall therefore decompose the $A$ fields in momentum modes (the time constraint and gauge constraint are here absorbed in the $\tilde{A}$ definition for brevity)
 \begin{equation}
 A^\mu(x) = \int d^³ x \left( \tilde{A^\mu}(k) e^{ik x} +  \tilde{(A^\mu)^*}(k)e^{-ik x}  \right)
 \end{equation}
 Assuming that regions defined by the same topological susceptibility are large 
 enough for field configurations to be approximately in this momentum eigenstate, one gets a thermal occupation number
 \begin{equation}
   f(p,q) = \exp\left[V_\theta \sum_i \frac{\hat{\theta}_i \tilde{Z}(p,q)}{T} \right]
 \end{equation}
 where
\begin{equation}
 \tilde{Z}(p,q) =\epsilon_{\alpha \beta \mu \nu} Z^{\alpha \beta}(p) Z^{\mu \nu} (q) \eqcomma Z^{\mu \nu}(p)= \epsilon^{\mu \nu \zeta \iota}  p_{\zeta} \tilde{A}_{ \iota}(p)
 \end{equation}
Note that $\hat{\theta}_i$ is a random variable that can be positive or negative, and the sum is taken over topological correlation lengths regions, each of volume $V_\theta$.   Both the variance $\ave{\theta^2}$ and $V_\theta$ are parameters that need to be computed from non-perturbative QCD.  Since this dynamics is non-perturbative, it is expected, and \cite{sharma,fodor} confirms, that $\ave{\theta^2}$ goes rapidly down above the deconfinement temperature (in Fig. S28 of \cite{fodor} $\ave{\theta^2} \sim \chi/T^4$).  $V_\theta$ should be $\sim 1$ fm in size but the quantitative value is not known.

Now, $\tilde{A}^{\gamma}(k)$ and $\epsilon^\mu(k,\alpha)$ are parallel for $\alpha=1$ and antiparallel for $\alpha=-1$.   I also consider that for an $\etaprime$ to form,spins of gluons have to point in opposite directions.
Hence,
following Eq. \ref{eq:vertex} then, the number of $\etaprime$ produced will have an abundance $\Delta n(p)$ w.r.t. equilibrium expectation values n(P) scaling as 
\begin{equation}
  \label{nptopo}
  \frac{\Delta n(P)}{n(p)} \sim \int d^3 p d^3q \abs{H(p^2, q^2, P^2) \tilde{Z}(p,q)}^2  \sum_i  \exp\left[ \frac{V_\theta}{T} \hat{\theta_i} \tilde{Z}(p,q) \right]
\end{equation}
The gluon momentum distributions $\tilde{A}^\mu(k)$ in $\tilde{Z}$ are either given by the same Juttner distributions of Eq. \ref{juttner}
\[\  \abs{\tilde{A}(p)}^2 \sim \int dx f_g(x,p) \]
or perhaps given by a glasma description \cite{glasma} (which should enhance topological configurations).   Note that both positive and negative values of $\hat{\theta}$ deviate from the equilibrium value.  This is consistent with the fact that $\etaprime$ is a pseudoscalar particle, insensitive to the actual value of spin.  It is only either when $\hat{\theta}=0$ or cancels out to zero that production is fully the one expected from thermal equilibrium.

Here, the main dependence driving $\etaprime$ abundance is the topological susceptibilty $\hat{\theta}/T$ and the number of regions $V_\theta$.  
Because of our complete lack of knowledge of the distribution of $\hat{\theta}$ and its correlation length in spacetime $V_\theta$ I feel I cannot give a quantitative estimate, even an elementary one like in the previous section, of Eq. \ref{nptopo}.   If one follows the ``chiral chemical potential'' ansatz  ($\mu_5$ with topological charge density $n_5$)  developed in \cite{magnetic1} and assumes $V_\theta \sim n_5^{-1}$ dimensional analysis would lead to a formula of the type
\begin{equation}
  \label{topeta}
  \frac{\etaprime}{\pi}  \sim \frac{\Delta^3 (m_{\eta}^{\prime})^2}{VT^2} K_2\left( \frac{m_\etaprime}{T} \right) \sum_{i=1}^{V n_5}  \exp \left[ \frac{\mu_5}{T} \hat{\theta_i}  \right]
  \end{equation}
where once again I have a non-extensive volume dependence due to the appearance of clusters with the same value of $\hat{\theta}$ and charge density $n_5$.   

Eq. \ref{topeta} allows to make some qualitative considerations.   \cite{fodor} suggests $\ave{\theta^2}$ remains approximately constant for $T=100-200$ MeV and falls very rapidly at higher temperatures.  Thus, at the temperature commonly associated with statistical hadronization the effects described in this section should be prevalent unless too many regions of $V_\theta$ size will cause topological configurations to cancel out in the $\sum_i \hat{\theta}_i$ of Eq. \ref{nptopo}.

Since the statistical model assumes the Hamiltonian expectation value to be given by the particle mass, without the inclusion of a term such as Eq.  \ref{deltah}, the most straight-forward prediction of relevance of chiral susceptibilities in particle production is a strong deviation of $\etaprime$ from the statistical model expectation value, perhaps for a characteristic momentum at which the integral in Eq. \ref{nptopo} becomes particularly peaked.

If $V_\theta$ is of the order of the nucleon size, this deviation should be prominent in small ($pp,pA$) systems but would go away for systems larger than the nucleus, i.e. central collsions, due to the cancellation in Eq. \ref{nptopo} of various $\hat{\theta}$ terms.   So far $\etaprime$ was compared to statistical models just in $e e$ collisions \cite{becastat}, where it appears to fit reasonably well, but not other systems.
Hence  even such a qualitative prediction is not open to verification without further experimental measurements.   Enhanced Event-by-event fluctuations of $\etaprime$ would be a more direct probe of $\ave{\theta^2}$, but, as clearly shown in the next section, it is most likely impossible to measure these in experiment.

Of course, coalescence should happen around hadronization temperature
the turnoff seen on the lattice at higher temperature should not, naively, impact on the $\etaprime$ abundance.
However, if the ``in-QGP'' production mechanism argued for in \cite{rene1,rene2} is applicable to the $\etaprime$, this particle becomes a very important probe of chiral susceptibility, althoguh in this case the loss term \cite{sangyong2} becomes important.
 \section{Discussion}
 The obvious observable here is the global abundance of $\etaprime$ compared to another particle, such as the $\pi$ or the $\phi$.  Which of these two should be the denominator in the ratio is a decision best left to experimentalists.
 The $\phi$ meson, through its dimuon decay, can be reconstructed with greater reliability than $\pi^0 \rightarrow \gamma \gamma$, does not have a background autocorrelation due to the same decay mode, and has a similar mass to the $\etaprime$.    However, it is also a much rarer particle than the $\pi^0$ and it also strongly depends on $\gamma_s$, which has a non-trivial dependence on centrality.  If the isospin chemical potential does not depend on centrality, something true for most experimentally accessible energies, $\etaprime/\pi^{+,-}$ provides
 a relatively reliable denominator free of autocorrelation.

 The signature for non-trivial dynamics is a variation of the ratio specifically on impact parameter.  Usually particle ratios are constant in the thermodynamic limit, strangeness seems to depend on multiplicity rather than initial geometry.   This can be experimentally probed either by comparing different system sizes same multiplicity bins, or by ''event engineering'', using event bins with similar event-by-event $v_2$.
 \begin{figure*}
   \begin{center}
\epsfig{width=1.\textwidth,figure=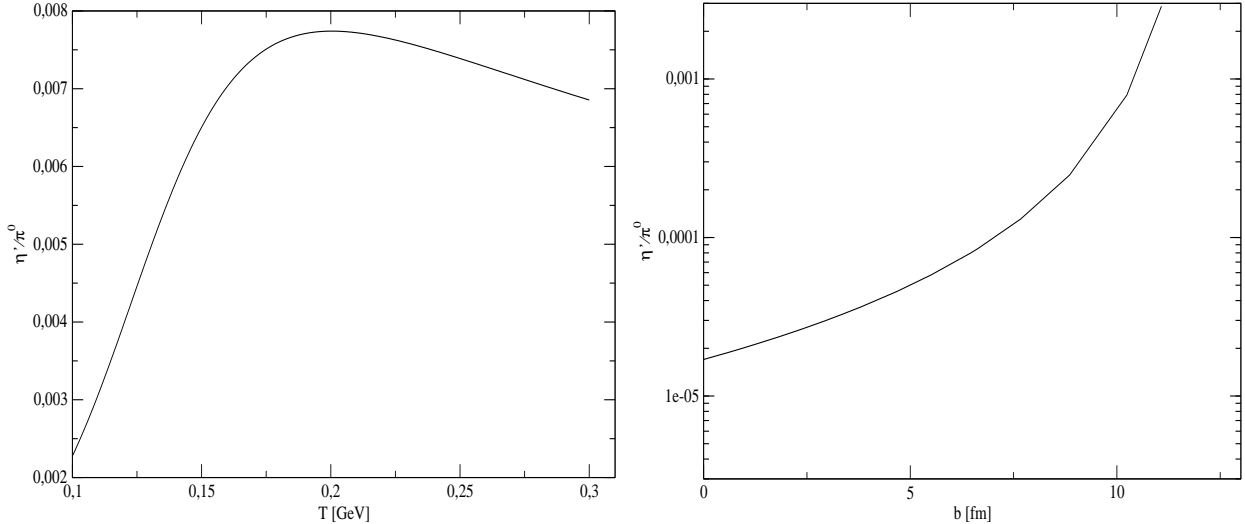}
\caption{\label{figeta}Predictions for $\etaprime/\pi$ dependence, assuming a zero isospin chemical potential.  The left panel (a) has a background statistical model estimate, temperature-driven and independent on volume and impact parameter, calculated in \cite{share}.  The right panel (b) a vorticity-driven calculation using the methods of section \ref{vortical}.}
\end{center}
\end{figure*}
 A strong obstacle in the analysis suggested in this work is the difficulty of reconstruction.  The only promising decay for the $\etaprime$ seems to be \cite{pdg} the $2.2\%$ $\etaprime \rightarrow \gamma \gamma$ decay, although perhaps the $\etaprime \rightarrow \rho^0 \gamma$ is achievable (its branching ratio is much larger, $30\%$).
 This is considerably larger than the $\phi \rightarrow \mu \mu$ decays ($\sim 10^{-4}$) observed by NA50 \cite{na50} and CERES \cite{ceres}
 The comparative abundance of the two particles, from thermal considerations, should be similar since the mass of the $\etaprime$ is of the order of the temperature around hadronization, which compensates of the $\phi$s higher degeneracy.

 That said, the $\pi^0 \rightarrow \gamma \gamma$ decays will provide a truly formidable background to this measurement.   However, an array of techniques can be used to separate this background from the signal.   The macroscopic lifetime of the $\pi^0$, the lower momentum of the $\pi^0\rightarrow \gamma \gamma$ decay products and the larger opening angle of this decay can all be used to distinguigh $\etaprime \rightarrow \gamma \gamma$ candidates from $\pi^0 \rightarrow \gamma \gamma$ background and the direct photon production from the collision.
 A detector with good electromagnetic calorimetry in the range of the $\etaprime$ decay kinematics, capable of estimating both the momentum and direction of each $\gamma$ with sufficient precision, should be able to reconstruct the $\etaprime$ peak given sufficient luminosity, unless the $\etaprime$ has been considerably modified in-medium.
 
 The obvious theoretical background to the effects examined here are other mechanisms of $\etaprime$ production.   Indeed, I interpreted the production as a kind of coalescence at hadronization, neglecting kinetic production and in-medium modification of the sort examined in \cite{sangyong2}.   To what extent this is a good approximation is highly dependent on how hadronization actually occurs.   If it occurs mostly through coalescence, and if symmetries distinguishing the $\etaprime$ from other particles are not altered by the medium, then the mechanism proposed here might be dominant since ``statistical'' coalescence from other quarks will be suppressed by parity conservation.  If these assumptions do not hold, or if the system is at chemical equilibrium around hadronization, the background will be that of the right panel of Fig. \ref{figeta}.
 
 This is why I would like to concentrate on the variation with centrality and system size rather than the absolute value when finding a signature.   As is well-known, the statistical model in the thermodynamic limit does not depend on volume \cite{pbm}.   A residual volume dependency on strangeness seems to scale with multiplicity but reach a plateau for high multiplicity AA collisions \cite{davidn}.    While a suppression has been seen in resonances in high multiplicity events \cite{markert}, explained either by rescattering \cite{markert} or initial state effects \cite{pythia,reso1,reso2}, this suppression seems to be specific to some short-lived Regge excitations which are longer lived but also strongly interact with the hadron gas (the $\Lambda(1520)$ and $K^*$ appear suppressed, but the $\Sigma^*$ and $\Xi^*$ do not \cite{reso1,reso2}).  The long-lived and non-strange $\etaprime$, with a comparatively weak in-medium cross-section \cite{referefexp2.5} will thus be less affected by hadronic in-medium rescatterings and will therefore conserve more memory of the production mechanism.
 
 Both of the mechanisms examined in the preceding two sections, when applied to high multiplicity collisions, go directly against these trends.   High multiplicity means high centrality events, where vorticity is expected to be lower.  The anti-correlation between vorticity and impact parameter, together with the yield of $\pi$ and $\phi$ linearly increasing with volume in the statistical model leads to the strong decrease with centrality seen in the right panel of Fig. \ref{figeta}.  No other phenomenon to the authors knowledge gives the same dependence for a particle ratio.

  On the other hand, a strong variation of $\etaprime/\pi$ with multiplicity reaching a statistical model plateau for more central events can be seen as an indication that topological fluctuations might be contributing to $\etaprime$ abundance.  A further disentanglement between multiplicity and impact parameter, possible with event engineering or simply system size scan (which can also disentangle rescattering from  initial state effects, which scale as the transverse multiplicity density), can shed light to the exact mechanism.  The estimates in Fig. \ref{figeta} show that it is the impact parameter dependence, rather than the absolute value, that need to be probed for new physics.

  The particle correlation studies presented in \cite{referefexp2,referefexp2.5} makes one hopeful a detectable contribution from chiral mechanisms to production of $\etaprime$ does indeed occur, but the methods used there can not easily isolate the source of centrality variation advocated in this paper from other dynamical mechanisms,present throghout the fireballs evolution, affecting pion correlations.   This is why a system size scan of $\etaprime$ ratios obtained from direct resonance reconstruction is both a promising probe of chiral physics and likely to yield a positive result.
  
 In conclusion, the impact parameter dependence of the $\etaprime/\pi$ ratio is a possible probe for the emergence of both vortical effects (transference between quark gluon plasma angular momentum and spin) and possibly the presence of topological excitations.    I hope experimentalists, especially those with precise electromagnetic calorimetry, will soon find if such an observable merits further phenomenological and theoretical investigation.

 GT acknowledges support from FAPESP proc. 2017/06508-7, partecipation in FAPESP tematico 2017/05685-2 and CNPQ bolsa de
 produtividade 301996/2014-8.  This work is a part
of the project INCT-FNA Proc. No. 464898/2014-5
 I wish to thank Sangyong Jeon and David Chinellato for discussions.

\end{document}